\documentclass[12pt]{article}
\usepackage[dvips]{color}
\usepackage{epsfig}
\usepackage{amsmath,amssymb}
\usepackage{graphicx}
\usepackage{cite}
\usepackage{subcaption}
\usepackage{hyperref}
\usepackage{bm}

\begin{document}

\title{\bf Investigating bounds on the extended uncertainty principle metric through astrophysical tests}
\author{{ \"{O}zg\"{u}r \"{O}kc\"{u} \thanks{Email: ozgur.okcu@ogr.iu.edu.tr}\hspace{1mm},
Ekrem Aydiner \thanks{Email: ekrem.aydiner@istanbul.edu.tr}} \\
{\small {\em Department of
		Physics, Faculty of Science, Istanbul University,
		}}\\{\small {\em Istanbul, 34134, Turkey}} }
\maketitle

\begin{abstract}
In this paper, we consider the gravitational tests for the extended uncertainty principle (EUP) metric, which is a large-scale quantum correction to Schwarzschild metric. We calculate gravitational redshift, geodetic precession, Shapiro time delay, precession of Mercury and S2 star's orbits. Using the results of experiments and observations, we obtain the lower bounds for the EUP fundamental length scale $L_{*}$. We obtain the  smallest bound $L_{*} \sim9\times 10^{-2}$m for gravitational redshift, and the largest bound $L_{*} \sim4\times 10^{10}$m for the precession of S2's orbit.\\

\end{abstract}

\section{Introduction}

\label{sec1}
	
\section{Introduction}
\label{intro}
Modifications of Heisenberg uncertainty principle (HUP) play a vital role in gravitational physics. There are two kinds of modifications. First kind of modification takes into account the quantum gravity effects near the Planck scale, and is called the generalized uncertainty principle (GUP). The simplest form of GUP is given by \cite{Maggiore1993}  
\begin{equation}
\label{SimpleGUP}
\Delta x\Delta p\leq1+\beta L_{Pl}^{2}\Delta p^{2},
\end{equation}
where $\beta $ is a dimensionless GUP parameter.\footnote{We use the units $\hbar=c=1$ through the paper. We only restore the physical constants for the numerical calculations.} Apart from GUP, a second kind of modification takes into account a long scale correction, and is called the extended uncertainty principle (EUP). The simplest form of EUP is given by \cite{Mureika2019}
\begin{equation}
\label{SimpleEUP}
\Delta x\Delta p\leq1+\frac{\alpha}{L_{*}^{2}}\Delta x^{2},
\end{equation}
where  $L_{*}$ is a fundamental length scale and $\alpha$ is a dimensionless EUP parameter.\footnote{Taking into account both momentum and position uncertainty corrections to HUP, a third kind of modification is also possible. It is called generalized extended uncertainty principle (GEUP), and is given by $\Delta x\Delta p\leq1+\beta L_{Pl}^{2}\Delta p^{2}+\frac{\alpha}{L_{*}^{2}}\Delta x^{2}$.}

Since GUP includes quantum gravity effects, it has been intensively studied in the literature. Various GUP models were proposed \cite{Maggiore1993,Scardigli1999,Das2008,Ali2009,Chang2018}. GUP may totally prevents the black hole evaporation. Therefore, black hole thermodynamics can be considered in the context of GUP \cite{Adler2001,Xiang2009,Scardigli2011,Nozari2012}. Investigations of GUP can be extended to different applications of cosmology \cite{Awad2014,Salah2017,Khodadi2018,Okcu2020}, deformed quantum and statistical mechanics\cite{Chang2018,Kempf1995,Nozari2012b,Vagenas2019}, etc.\footnote{The literature on GUP is comprehensive. Interested reader may refer to review in ref. \cite{Tawfik2014}}

On the other hand, EUP affects large scale gravitational physics since it includes quantum effects at large distance. Recently, much attention has been focused on EUP. In ref. \cite{Bambi2008}, Bambi and Urban derived EUP from a gedanken experiment in de Sitter spacetime. Another derivation, which based on modified commutation relation from a non-Euclidean space, can be found in ref.\cite{Filho2016}. A new type of EUP was proposed in ref. \cite{Chung2019}. The author studied the deformations of classical mechanics, calculus, and quantum mechanics for the new type of EUP. Just like GUP, EUP also gives some interesting results for the modification of black hole thermodynamics and Friedmann equations. In ref. \cite{Dabrowski2019}, Dabrowski and Wagner obtained EUP relations for Rindler and Friedmann horizons. They studied black hole temperature and entropy for both relations. They showed that temperature decreases while entropy increases. In ref. \cite{Moradpour2019}, Moradpour et. al. interestingly showed that the EUP correction to black hole entropy is similar to Rényi entropy. Considering the Bohr like approach, they also studied the stable-unstable phase transition for an excited black hole. In another paper \cite{Chung2019b}, Chung and Hassanabadi studied Schwarzschild black hole thermodynamics and Unruh effect for EUP. Unlike GUP case, they found a lower bound for the black hole temperature. They also showed that Unruh temperature increases for the EUP correction. In ref. \cite{Gine2020}, Giné and Luciano obtained the modified inertia for two EUP relations. They showed that EUP may provide a natural explanation for MoND. EUP can also be considered for thermodynamics of  Friedmann-Robertson-Walker (FRW) universe. For example, Zhu et. al. studied Friedmann equations for GUP and EUP \cite{Zhu2009}. They found corrected entropy of apparent horizon for GUP and EUP. They obtained modified Friedmann equations from modified entropy and first law of thermodynamics at apparent horizon. It is also possible to consider EUP corrections for conventional thermodynamics systems. In ref. \cite{Chung2019c}, EUP modified number of microstates was obtained to investigate the thermodynamics of monatomic and interacting gas models.

Besides the above mentioned studies, EUP may modify the black hole solutions. EUP black hole solution was proposed by Mureika \cite{Mureika2019}. He obtained the modified black hole characteristics such as horizon radius, ISCO, and  photosphere. It was shown that if $L_{*}$ is $10^{12}-10^{14}$ m, EUP will become relevant for the black holes in the range $10^{9}-10^{11} M_{\odot}$. Finally, he calculated the Hawking temperature of EUP black hole, and found that EUP black hole temperature has smaller than standard temperature. Recently, EUP black holes considered for gravitational lensing \cite{Lu2019}, shadow and weak deflection angle \cite{Kumaran2020,Panting2021}. 

In this paper, we would like to find lower bounds of new fundamental length scale $L_{*}$. Therefore, we will study some astrophysical tests such as gravitational redshift, geodetic precession, Shapiro time delay, precession of Mercury and S2 star's orbits for EUP metric \footnote{Astrophysical tests may provide constraints for various modifed theories of gravity. The reader may refer to refs. \cite{Ali2015,Khodadi2017,Bahamonde2020,Zhu2020}}.  Getting constraints on $L_{*}$ may provide us a better understanding of large scale EUP effects. Besides, finding bounds on EUP from experiments and observations is sparse in the literature. In the  GUP case, the studies on this direction are not new.  There are a lot of studies amied to obtain upper bounds from various experiments and observations \cite{Das2008,Ali2011,Das2011,Pedram2011,Marin2013,Gosh2014,Bawaj2015,Scardigli2015,Ali2015b,Khodadi2015,Gao2016,Gao2017,Feng2017,Kouwn2018,Bushev2019,Khodadi2019,Neves2020,Giardino2021,Das2021,Fellepa2021,Aghababaei2021}. As for EUP case, constraints on EUP were studied in refs. \cite{Lu2019,Nozari2019,Aghababaei2021,Illuminati2021}. In ref. \cite{Lu2019}, Lu and Xie obtained constraints on $L_{*}$ from gravitational lensing. In ref.\cite{Aghababaei2021}, Aghababaei et. al. set bounds on GUP and EUP from Hubble tension. In ref. \cite{Nozari2019}, Nozari and Dehghani found bounds on EUP for both Newtonian and relativistic cosmologies based on Verlinde's entropic gravity. In ref. \cite{Illuminati2021}, assuming equality between EUP and gravity sector of Standart Model Extension modified Hawking temperatures, Illuminati et. al. found bounds on EUP dimensionless parameters.

The rest of  paper is arranged as follows. In the next section, we briefly review the EUP metric and derive effective potential of a particle around orbit in EUP metric. In the third section, we use the EUP metric to compute gravitational redshift, geodetic precession, Shapiro time delay, precession of Mercury and S2 star's orbits.  Finally, we discuss our results.

\section{The extended uncertainty principle metric}
\label{sec2}
In this section, we review the EUP metric proposed in ref. \cite{Mureika2019}. Considering the confinement of N gravitons to Schwarzschild radius $\Delta X\sim r_{S}=2G_{N}M$, each  graviton momentum uncertainty $\Delta p_{g}$ is given by \cite{Mureika2019}
\begin{equation}
\label{momUnceOfGra}
\Delta p_{g}\sim\frac{1}{2G_{N}M}\left(1+\frac{4\alpha G_{N}^{2}M^{2}}{L_{*}^{2}}\right),
\end{equation}
where $M$ is the black hole mass. If the total mass of N gravitons is considered, then we have $\frac{N}{2G_{N}M}$. Therefore, total momentum uncertainty $\Delta P$ is given by
\begin{equation}
\label{TotMomUnce}
\Delta P\sim M\left(1+\frac{4\alpha G_{N}^{2}M^{2}}{L_{*}^{2}}\right).
\end{equation}
One may interpret eq. (\ref{TotMomUnce}) as EUP corrected mass
\begin{equation}
\label{EUPCorrectedMass}
M_{EUP}=M\left(1+\frac{4\alpha G_{N}^{2}M^{2}}{L_{*}^{2}}\right),
\end{equation}
and assume that EUP correction corresponds to the stress-energy tensor,
\begin{equation}
\label{EUPStEnTen}
M_{EUP}=\int d^{3}x\sqrt{g}\left(T_{0GR}^{0}+T_{0EUP}^{0}\right).
\end{equation}
Replacing  $M$ with $M_{EUP}$ leads to EUP corrected Schwarzschild metric, i.e, 
\begin{equation}
\label{EUPMetric}
F(r)=1-\frac{2G_{N}M}{r}\left(1+\frac{4\alpha G_{N}^{2}M^{2}}{L_{*}^{2}}\right),
\end{equation}
and the event horizon of EUP metric is given by
\begin{equation}
r_{H}=2G_{N}M\left(1+\frac{4\alpha G_{N}^{2}M^{2}}{L_{*}^{2}}\right).
\end{equation}
At this point, we give some comments on dimensionless EUP parameter $\alpha$. It is assumed that $\alpha$ is taken to be order of unity. So, we only get bounds on new fundamental length scale $L_{*}$. Choosing $\alpha=-1$ seems problematic. If $\alpha$ is negative, there is a maximum mass for $r_{H}=0$. Another problem arises as repulsive potential for sufficiently large masses. (Please see ref. \cite{Mureika2019} for more details.) Therefore, we exclude the negativity of $\alpha$, and consider $\alpha=1$.

\subsection{Particle motion in the EUP metric}
\label{sec2.1}
We begin to consider particle in equatorial plane $\theta=\pi/2$. We  give the Lagrangian of particle \cite{Bambi2018},
\begin{equation}
\label{Lagra}
\mathcal{L}=\frac{1}{2}g_{\mu\nu}\dot{x}^{\mu}\dot{x}^{\nu}=\frac{1}{2}\left[-F(r)\dot{t}^{2}+\frac{\dot{r}^{2}}{F(r)}+r^{2}\dot{\phi}^{2}\right],
\end{equation}
where $\dot{x}^{\mu}=dx^{\mu}/d\lambda$, and $\lambda$ is the affine parameter. Following the standard procedure, constants of motion can be obtained
\begin{equation}
\label{Energy}
p_{t}=\frac{\partial\mathcal{L}}{\partial\dot{t}}=-F(r)\dot{t}=-e\quad\Longrightarrow\quad\dot{t}=\frac{e}{F(r)},
\end{equation}
\begin{equation}
\label{AnguMom}
p_{\phi}=\frac{\partial\mathcal{L}}{\partial\dot{\phi}}=r^{2}\dot{\phi}=\ensuremath{\ell}\quad\Longrightarrow\quad\dot{\phi}=\frac{\ell}{r^{2}},
\end{equation}
where $e$ and $\ell$ denote the energy and angular momentum of the particle, respectively. Employing above expressions in $g_{\mu\nu}\dot{x}^{\mu}\dot{x}^{\nu}=-k$ ($k=0$ for masseles particle and $k=1$ for massive particle), we find
\begin{equation}
\label{beforeVeffForm}
-\frac{e^{2}}{F(r)}+\frac{\dot{r}^{2}}{F(r)}+\frac{\ell^{2}}{r^{2}}=-k.
\end{equation}
Using eq. (\ref{EUPMetric}), the above expression can be rearranged as
\begin{equation}
\label{VeffForm}
\frac{e^{2}-k}{2}=\frac{1}{2}\dot{r}^{2}+V_{eff},
\end{equation}
where the effective potential $V_{eff}$ is given by
\begin{equation}
\label{Veff}
V_{eff}=-k\frac{G_{N}M}{r}+\frac{\ell^{2}}{2r^{2}}-\frac{G_{N}M\ell^{2}}{r^{3}}-\alpha\frac{4G_{N}^{3}M^{3}}{L_{*}^{2}r}\left(k+\frac{\ell^{2}}{r^{2}}\right).
\end{equation}

\section{Astrophysical tests of the EUP metric}
\label{sec3}
In this section, we focus on gravitational tests of EUP metric. Comparing our results with observations and experiments, we find bounds for fundamental length scale $L_{*}$.

\subsection{Gravitational redshift}
\label{sec3.1}
Let us first consider the gravitational redshift of electromagnetic signal. If the electromagnetic signal travels from point A to point B in a gravitational field, then gravitational redshift is defined by \cite{Bambi2018}
\begin{equation}
\label{GRedShift}
\frac{\nu_{B}}{\nu_{A}}=\sqrt{\frac{F(r_{A})}{F(r_{B})}}.
\end{equation}
For EUP metric in eq. (\ref{EUPMetric}), the above expression is given by
\begin{equation}
\label{GRedShift2}
\frac{\nu_{B}}{\nu_{A}}=\sqrt{\frac{1-\frac{2G_{N}M}{r_{A}}\left(1+\frac{4\alpha G_{N}^{2}M^{2}}{L_{*}^{2}}\right)}{1-\frac{2G_{N}M}{r_{B}}\left(1+\frac{4\alpha G_{N}^{2}M^{2}}{L_{*}^{2}}\right)}}.
\end{equation}
Expanding eq. (\ref{GRedShift}), the frequency shift is given by
\begin{eqnarray}
\frac{\Delta\nu}{\nu_{A}}=\frac{G_{N}M(r_{A}-r_{B})}{r_{A}r_{B}}\left[1+\frac{G_{N}M\left(3r_{A}+r_{B}\right)}{2r_{A}r_{B}}+\frac{4\alpha G_{N}^{2}M^{2}}{L_{*}^{2}}\left(1+\frac{G_{N}(3r_{A}+r_{B})}{r_{A}r_{B}}\right)\right],
\label{freqShift}
\end{eqnarray}
where $\Delta \nu=\nu_{B}-\nu_{A}$. 

In order to get a bound for $L_{*}$, we refer to Pound-Snider experiment \cite{Pound1965} which was carried out in a tower with height $h=22.86$ m. Relative deviation of frequency is
\begin{equation}
\label{relDevFreq}
\frac{\frac{\Delta\nu}{\nu_{A}}-\left(\frac{\Delta\nu}{\nu_{A}}\right)^{GR}}{\left(\frac{\Delta\nu}{\nu_{A}}\right)^{GR}}<0.01.
\end{equation}
Using eq. (\ref{freqShift}) in eq. (\ref{relDevFreq}) yields
\begin{eqnarray}
\label{GRedEUP1}
\frac{\alpha}{L_{*}^{2}}<\frac{c^{4}}{4G_{N}^{2}M^{2}}\left(\frac{1}{100}-\frac{G_{N}M(3r_{A}+r_{B})}{2r_{A}r_{B}c^{2}}\right)\left(1+\frac{G_{N}M(3r_{A}+r_{B})}{c^{2}r_{A}r_{B}}\right)^{-1},
\end{eqnarray}
where $M=M_{\oplus}=5.972\times10^{24}$kg, $R_{A}=R_{\oplus}=6378$km, and $R_{B}=R_{\oplus}+h$. The lower bound of $L_{*}$ is given by
\begin{equation}
\label{GRedBound2}
9\times10^{-2}m\lesssim L_{*}.
\end{equation}

\subsection{Geodetic precession}
\label{sec3.2}
Let us consider a gyroscope rotating in an orbit around a spherical massive body. General relativity predicts that the spin direction of gyroscope changes. This phenomena is called geodetic precession. A gyroscope with a spin four-vector $\boldsymbol{s}$ is characterized by \cite{Hartle2014}
\begin{equation}
\label{gyroEq}
\frac{ds^{\alpha}}{d\tau}+\Gamma_{\mu\nu}^{\alpha}s^{\mu}u^{\nu}=0,
\end{equation}
where $\Gamma_{\mu\nu}^{\alpha}$ is Christoffel symbol. We call eq. (\ref{gyroEq}) gyroscope equation. It determines the components of spin vector. The spin four-vector $\boldsymbol{s}$ and velocity four-vector $\boldsymbol{u}$ satisfy the following conditions 
\begin{equation}
\label{FirstCond}
\boldsymbol{s}.\boldsymbol{u}=g_{\mu\nu}s^{\mu}u^{\nu}=0,\qquad \boldsymbol{s}.\boldsymbol{s}=g_{\mu\nu}s^{\mu}s^{\nu}=s_{*}^{2},
\end{equation}
where $s_{*}$ is the magnitude of spin. Choosing equatorial plane ($\theta=\pi/2$) and circular orbit ($\dot{r}=0=\dot{\theta}$) obviously simplifies the problem. The components of velocity four-vector are given by
\begin{equation}
\label{fourVeloVector}
\boldsymbol{u}=u^{t}(1,0,0,\Omega),
\end{equation}
where $\Omega=d\phi/dt$ is the orbital angular velocity. Since $\dot{r}$ vanishes for the stable circular orbits, eq. (\ref{VeffForm}) yields
\begin{equation}
\label{ForStableCircularOrbit1}
\frac{e^{2}-1}{2}=V_{eff},
\end{equation}
and circular orbit radius $R$ is found from
\begin{equation}
\label{ForStableCircularOrbit2}
\frac{dV_{eff}}{dr}=0.
\end{equation}
From eqs. (\ref{ForStableCircularOrbit1}) and (\ref{ForStableCircularOrbit2}), one gets
\begin{eqnarray}
\label{e2}
e^{2}=\left[1-\frac{2G_{N}M}{R}\left(1+\frac{4\alpha G_{N}^{2}M^{2}}{L_{*}^{2}}\right)\right]^{2}\left[1-\frac{3G_{N}M}{R}\left(1+\frac{4\alpha G_{N}^{2}M^{2}}{L_{*}^{2}}\right)\right]^{-1},
\end{eqnarray}
\begin{equation}
\label{l2}
\ell^{2}=G_{N}MR\left(1+\frac{4\alpha G_{N}^{2}M^{2}}{L_{*}^{2}}\right)\left[1-\frac{3G_{N}M}{R}\left(1+\frac{4\alpha G_{N}^{2}M^{2}}{L_{*}^{2}}\right)\right]^{-1},
\end{equation}
\begin{equation}
\label{angularVelo}
\Omega=\frac{d\phi}{d\tau}\frac{d\tau}{dt}=\frac{F(R)}{R^{2}}\frac{\ell}{e}=\sqrt{\frac{G_{N}M}{R^{3}}\left(1+\frac{4\alpha G_{N}^{2}M^{2}}{L_{*}^{2}}\right)}.
\end{equation}

Now, let us begin to solve the gyroscope equations. We suppose that $\boldsymbol{s}$ is radial directed at the beginning, i.e., only $s^{r}(0)\neq0$. From orthogonality condition in eq. (\ref{FirstCond}), the relation between components $s^{t}$ and $s^{\phi}$ is given by
\begin{equation}
\label{sTsPhiRel}
s^{t}=\Omega R^{2}\left[1-\frac{2G_{N}M}{R}\left(1+\frac{4\alpha G_{N}^{2}M^{2}}{L_{*}^{2}}\right)\right]^{-1}s^{\phi}.
\end{equation}
From eqs. (\ref{fourVeloVector}) and (\ref{sTsPhiRel}), the gyroscope equations are given by
\begin{equation}
\label{radialGE}
\frac{ds^{r}}{d\tau}+\Omega\left[3G_{N}M\left(1+\frac{4\alpha G_{N}^{2}M^{2}}{L_{*}^{2}}\right)-R\right]s^{\phi}u^{t}=0,
\end{equation}
\begin{equation}
\label{thetaGE}
\frac{ds^{\theta}}{d\tau}=0,
\end{equation}
\begin{equation}
\label{phiGE}
\frac{ds^{\phi}}{d\tau}+\frac{\Omega}{R}s^{r}u^{t}=0.
\end{equation}
It is clearly seen that $s^{\theta}$ remains zero due to $s^{\theta}(0)=0$. Since $u^{t}=dt/d\tau$, eqs. (\ref{radialGE}) and (\ref{phiGE}) can be rearranged as
\begin{equation}
\label{radialGE2}
\frac{ds^{r}}{dt}+\left[3G_{N}M\left(1+\frac{4\alpha G_{N}^{2}M^{2}}{L_{*}^{2}}\right)-R\right]\Omega s^{\phi}=0,
\end{equation}
\begin{equation}
\label{phiGE2}
\frac{ds^{\phi}}{dt}+\frac{\Omega}{R}s^{r}=0,
\end{equation}
respectively. Substituting eq. (\ref{phiGE2}) into eq. (\ref{radialGE2}) leads to a second-order differential equation,
\begin{equation}
\label{phiGE3}
\frac{d^{2}s^{\phi}}{dt^{2}}+\tilde{\Omega}^{2}s^{\phi}=0,
\end{equation}
where $\tilde{\Omega}$ is defined by
\begin{equation}
\label{tildeOmega}
\tilde{\Omega}=\sqrt{1-\frac{3G_{N}M}{R}\left(1+\frac{4\alpha G_{N}^{2}M^{2}}{L_{*}^{2}}\right)}\Omega .
\end{equation}
One can solve the eqs. (\ref{radialGE2}) and (\ref{phiGE3}) which give the results
\begin{equation}
\label{sR}
s^{r}=s_{*}\sqrt{1-\frac{2G_{N}M}{R}\left(1+\frac{4\alpha G_{N}^{2}M^{2}}{L_{*}^{2}}\right)}\cos\left(\tilde{\Omega}t\right),
\end{equation}
\begin{equation}
\label{sPh}
s^{\phi}=-s_{*}\frac{\Omega}{\tilde{\Omega}R}\sqrt{1-\frac{2G_{N}M}{R}\left(1+\frac{4\alpha G_{N}^{2}M^{2}}{L_{*}^{2}}\right)}\sin\left(\tilde{\Omega}t\right),
\end{equation}
where we employ the conditions $\boldsymbol{s}.\boldsymbol{s}=s_{*}^{2}$ and $s^{t}(0)=s^{\phi}(0)=0$.

The spin initially starts along a unit vector $\boldsymbol{e}_{\hat{r}}$. After one complete rotation in a time $P=2\pi/\Omega$, the change of spin direction is given by
\begin{equation}
\label{chanOfSpinDir}
\left[\frac{\boldsymbol{s}}{s_{*}}.\boldsymbol{e}_{\hat{r}}\right]_{t=P}=\cos\left(\frac{2\pi\tilde{\Omega}}{\Omega}\right).
\end{equation}
Therefore, the geodetic precession angle is given by
\begin{equation}
\label{geoPrAng}
\Delta\Phi_{geodetic}=2\pi-2\pi\sqrt{1-\frac{3G_{N}M}{R}\left(1+\frac{4\alpha G_{N}^{2}M^{2}}{L_{*}^{2}}\right)},
\end{equation}
which can approximately be written as
\begin{equation}
\label{geoPrAng2}
\Delta\Phi_{geodetic}\approx\Delta\Phi_{GR}\left(1+\frac{4\alpha G_{N}^{2}M^{2}}{c^{4}L_{*}^{2}}\right),
\end{equation}
where $\Delta\Phi_{GR}=\frac{3\pi G_{N}M}{Rc^{2}}$ is predicted by general relativity. 

In order to get a bound for $L_{*}$, we refer to measurements of Gravity Probe B (GPB) \cite{Everitt2011}, which was a satellite in a orbit around the Earth. Considering GPB was located at $642$km altitude and had $97.65$ min orbital period, the general relativity predicts $\Delta\Phi_{GR}=6606.1$mas/year. The measurement of GPB is given by
\begin{equation}
\label{GPBMeas}
\Delta\Phi_{geodetic}=(6601.8\pm18.3)mas/year,
\end{equation}
which gives $6620.1$mas/year and $6583.5$mas/year. Since later value imposes $\alpha=-1$, we consider maximum value, i.e., $6620.1$mas/year. Therefore, we found
\begin{equation}
2\times10^{-1}m\lesssim L_{*}.
\end{equation}
Up to now, we have considered Earth based experiments to constrain $L_{*}$. In the rest of paper, we consider gravitational tests for solar system and beyond.

\subsection{Shapiro time delay}
\label{sec3.3}
If an electromagnetic signal travels in a gravitational field, the travel time of signal takes longer than the travel time of the same signal in flat spacetime. This effect is called Shapiro time delay \cite{Shapiro1964}. In this section, we follow the arguments of ref. \cite{Bambi2018}.

Let us consider that the electromagnetic signal travels from a point $A$ to point $B$ in the Solar system. Without loss of generality, we again consider  the equatorial plane, i.e., $\theta=\pi/2$. Employing
\begin{equation}
\label{withHelp}
\frac{dr}{d\lambda}=\frac{dr}{dt}\frac{dt}{d\lambda}=\frac{dr}{dt}\frac{e}{F(r)},
\end{equation}
eq. (\ref{beforeVeffForm}) can be rearranged as
\begin{equation}
\label{beforeVeffForm2}
\frac{e^{2}}{F(r)^{3}}\left(\frac{dr}{dt}\right)^{2}+\frac{\ell^{2}}{r^{2}}-\frac{e^{2}}{F(r)}=0,
\end{equation}
for massless particles. For $r=r_{O}$ (the closest distance to Sun), one gets 
\begin{equation}
\label{intStep1}
\ell^{2}=\frac{er_{O}^{2}}{F(r_{O})}.
\end{equation}
Employing eq. (\ref{intStep1}) in eq. (\ref{beforeVeffForm2}), we find
\begin{equation}
\label{intStep2}
dt=\pm\frac{dr}{\sqrt{F(r)^{2}\left(1-\frac{F(r)r_{O}^{2}}{F(r_{O})r^{2}}\right)}}.
\end{equation}
Expanding in $r_{S}/r$ and $r_{S}/r_{O}$, eq. (\ref{intStep2}) can ben given in the integral form as follows:
\begin{eqnarray}
\label{timDelIntForm1}
&t=\int\frac{dr}{\sqrt{F(r)^{2}\left(1-\frac{F(r)r_{O}^{2}}{F(r_{O})r^{2}}\right)}}\approx\int\frac{rdr}{\sqrt{r^{2}-r_{O}^{2}}}+\int\left(1+\frac{r_{S}^{2}}{L_{*}^{2}}\alpha\right)\nonumber\\&\times\left(\frac{r^{2}r_{S}}{(r^{2}-r_{O}^{2})^{3/2}}+\frac{rr_{O}r_{S}}{2(r^{2}-r_{O}^{2})^{3/2}}-\frac{3r_{O}^{2}r_{S}}{2(r^{2}-r_{O}^{2})^{3/2}}\right)dr
\end{eqnarray}
So, we find the the travel times from  point $A$ to point $O$ and point $O$ to point $B$
\begin{eqnarray}
\label{travelTimeAC}
&t_{AO}=\sqrt{r_{A}^{2}-r_{O}^{2}}+\left(1+\frac{r_{S}^{2}}{L_{*}^{2}}\alpha\right)\left(\frac{r_{S}}{2}\sqrt{\frac{r_{A}-r_{O}}{r_{A}+r_{O}}}+r_{S}\ln\left(\frac{r_{A}+\sqrt{^{r_{A}^{2}}-r_{O}^{2}}}{r_{O}}\right)\right),
\end{eqnarray}
\begin{eqnarray}
\label{travelTimeCB}
&t_{BO}=\sqrt{r_{B}^{2}-r_{O}^{2}}+\left(1+\frac{r_{S}^{2}}{L_{*}^{2}}\alpha\right)\left(\frac{r_{S}}{2}\sqrt{\frac{r_{B}-r_{O}}{r_{B}+r_{O}}}+r_{S}\ln\left(\frac{r_{B}+\sqrt{^{r_{B}^{2}}-r_{O}^{2}}}{r_{O}}\right)\right),
\end{eqnarray}
respectively. The total travel time of signal is given by
\begin{equation}
\label{totalTravelTime1}
t_{tot}=2\left(t_{AO}+t_{BO}\right).
\end{equation}
For flat spacetime, it is given by
\begin{equation}
\label{totalTravelTime2}
\tilde{t}_{tot}=2\left(\sqrt{r_{A}^{2}-r_{O}^{2}}+\sqrt{r_{B}^{2}-r_{O}^{2}}\right).
\end{equation}
Considering $r_{O}\ll r_{A},r_{B}$, the time delay is given by
\begin{equation}
\label{timeDelay}
\delta t=t_{tot}-\tilde{t}_{tot}=4G_{N}M\left(1+\frac{4\alpha G_{N}^{2}M^{2}}{L_{*}^{2}}\right)\left(1+\ln\left(\frac{4r_{A}r_{B}}{r_{O}^{2}}\right)\right).
\end{equation}
In order to get a bound on $L_{*}$, we compare eq. (\ref{timeDelay}) with the time delay which is defined in parameterized Post-Newtonian (PPN) formalism \cite{Will2014}
\begin{equation}
\label{timeDelayPPN}
\delta t_{PPN}=4G_{N}M\left(1+\left(\frac{1+\gamma}{2}\right)\ln\left(\frac{4r_{A}r_{B}}{r_{O}^{2}}\right)\right),
\end{equation}
where $\gamma$ is a dimensionless PPN parameter. We refer to measurements of Cassini spacecraft \cite{Bertotti2003}. The constraint on $\gamma$ is $|\gamma-1|<2.3\times10^{-5}$.
Comparing eqs. (\ref{timeDelay}) with (\ref{timeDelayPPN}), we get
\begin{equation}
\label{TdEqTdPPN}
\frac{8\alpha G_{N}^{2}M^{2}}{c^{4}L_{*}^{2}}\left(1+\frac{1}{\ln\left(\frac{4r_{A}r_{B}}{r_{O}^{2}}\right)}\right)=|\gamma-1|<2.3\times10^{-5}.
\end{equation}
Finally, taking $r_{A}=1$AB, $r_{B}=8.46$AB and $r_{O}=1.6R_{\odot}$, one gets
\begin{equation}
\label{TDBound}
9\times10^{5}m\lesssim L_{*}.
\end{equation}

\subsection{Precession of Mercury and S2 star's orbits}
\label{sec3.4}
Now let us turn our attention to the perihelion shift of Mercury and precession of S2's orbit. In this section, we follow the arguments of ref. \cite{Weinberg1972}. For a massive particle ($k=1$), eq. (\ref{beforeVeffForm}) can be rearranged as
\begin{equation}
\label{rDot}
\dot{r}=\pm\sqrt{e^{2}-F(r)\left(1+\frac{\ell^{2}}{r^{2}}\right)}.
\end{equation}
Dividing eq. (\ref{AnguMom}) by eq. (\ref{rDot}), we have
\begin{equation}
\label{dphi/dr}
\frac{d\phi}{dr}=\pm\frac{\ell}{r^{2}}\left[e^{2}-F(r)\left(1+\frac{\ell^{2}}{r^{2}}\right)\right]^{-1/2}.
\end{equation}
From eq. (\ref{dphi/dr}), one may write the orbital precession as
\begin{equation}
\label{totPrec}
\psi_{prec}=2\int_{r_{-}}^{r_{+}}\frac{\ell}{r^{2}}\left[e^{2}-F(r)\left(1+\frac{\ell^{2}}{r^{2}}\right)\right]^{-1/2}dr-2\pi,
\end{equation}
where $r_{+}$ and $r_{-}$ are the maximum and minimum points, respectively. Since $dr/d\phi$ vanishes for $r=r_{\pm}$, eq. (\ref{dphi/dr}) gives
\begin{equation}
\label{doubleEqs}
\frac{1}{r_{\pm}^{2}}+\frac{1}{\ell^{2}}=\frac{e^{2}}{\ell^{2}F(r_{\pm})}.
\end{equation}
Solving these equations yields
\begin{equation}
\label{e22}
e^{2}=\frac{F(r_{+})F(r_{-})(r_{+}^{2}-r_{-}^{2})}{r_{+}^{2}F(r_{-})-r_{-}^{2}F(r_{+})},
\end{equation}
\begin{equation}
\label{l22}
\ell^{2}=\frac{r_{+}^{2}r_{-}^{2}(F(r_{-})-F(r_{+}))}{r_{-}^{2}F(r_{+})-r_{+}^{2}F(r_{-})}.
\end{equation}
Substituting eqs. (\ref{e22}) and (\ref{l22}) into the integral in eq. (\ref{totPrec}), we have
\begin{equation}
\label{totPrec2}
\psi_{prec}=2\int_{r_{-}}^{r_{+}}\zeta^{-1/2}\frac{dr}{\sqrt{F(r)}r^{2}}-2\pi,
\end{equation}
where $\zeta$ is defined by
\begin{eqnarray}
\label{bracketsWithC}
\zeta=\frac{r_{-}^{2}\left(\frac{1}{F(r)}-\frac{1}{F(r_{-})}\right)-r_{+}^{2}\left(\frac{1}{F(r)}-\frac{1}{F(r_{+})}\right)}{r_{+}^{2}r_{-}^{2}\left(\frac{1}{F(r_{+})}-\frac{1}{F(r_{-})}\right)}-\frac{1}{r^{2}}=C\left(\frac{1}{r_{-}}-\frac{1}{r}\right)\left(\frac{1}{r}-\frac{1}{r_{+}}\right).
\end{eqnarray}
Since $\zeta$ vanishes for $r=r_{\pm}$, it can be expressed with the second line in above equation and the constant $C$ can be obtained in the limit $r\rightarrow\infty$. It is given by
\begin{eqnarray}
\label{CConst}
C=\frac{r_{-}^{2}F(r_{+})(F(r_{-})-1)-r_{+}^{2}F(r_{-})(F(r_{+})-1)}{r_{+}r_{-}(F(r_{+})-F(r_{-}))}=&1-2G_{N}M\left(1+\frac{4\alpha G_{N}^{2}M^{2}}{L_{*}^{2}}\right)\left(\frac{1}{r_{-}}+\frac{1}{r_{+}}\right),
\end{eqnarray}
or we can approximately write
\begin{equation}
\label{CAprrox}
C^{-1/2}\approx1+G_{N}M\left(1+\frac{4\alpha G_{N}^{2}M^{2}}{L_{*}^{2}}\right)\left(\frac{1}{r_{-}}+\frac{1}{r_{+}}\right).
\end{equation}
Therefore, total precession is given by
\begin{equation}
\label{totPrec3}
\psi_{prec}=2C^{-1/2}\int_{r_{-}}^{r_{+}}\left(\frac{1}{r_{-}}-\frac{1}{r}\right)^{-1/2}\left(\frac{1}{r}-\frac{1}{r_{+}}\right)^{-1/2}\frac{dr}{r^{2}\sqrt{F(r)}}-2\pi.
\end{equation}
This integral can be solved by choosing a suitable change of variable. So, we introduce
\begin{equation}
\label{ChangeOfVari}
\frac{1}{r}=\frac{1}{2}\left(\frac{1}{r_{+}}+\frac{1}{r_{-}}\right)+\frac{1}{2}\left(\frac{1}{r_{+}}-\frac{1}{r_{-}}\right)\sin\rho.
\end{equation}

For eq. (\ref{ChangeOfVari}), the integral in eq. (\ref{totPrec3}) is given 
\begin{eqnarray}
\label{totalPrecRho}
&\psi_{prec}=2\left[1+G_{N}M\left(\frac{1}{r_{-}}+\frac{1}{r_{+}}\right)\left(1+\frac{4\alpha G_{N}^{2}M^{2}}{L_{*}^{2}}\right)\right]\nonumber\\&\times\int_{-\pi/2}^{\pi/2} 1+\frac{G_{N}M}{2}\left[\left(\frac{1}{r_{-}}+\frac{1}{r_{+}}\right)+\left(\frac{1}{r_{+}}-\frac{1}{r_{-}}\right)\sin\rho\right]\left(1+\frac{4\alpha G_{N}^{2}M^{2}}{L_{*}^{2}}\right) d\rho-2\pi.
\end{eqnarray}
Finally, total precession is 
\begin{equation}
\label{finalTotalPrec}
\psi_{prec}=3\pi G_{N}M\left(1+\frac{4\alpha G_{N}^{2}M^{2}}{L_{*}^{2}}\right)\left(\frac{1}{r_{+}}+\frac{1}{r_{-}}\right),
\end{equation}
or 
\begin{equation}
\label{finalTotalPrec2}
\psi_{prec}=\frac{6\pi G_{N}M}{L}\left(1+\frac{4\alpha G_{N}^{2}M^{2}}{L_{*}^{2}}\right),
\end{equation}
where we use the semilatus rectum $L$ which is defined by 
\begin{equation}
\label{semilatusRectum}
\frac{1}{L}=\frac{1}{2}\left(\frac{1}{r_{+}}+\frac{1}{r_{-}}\right).
\end{equation}
In order to find a bound on $L_{*}$, we consider total precession in PPN formalism, which is given by \cite{Will2014}
\begin{equation}
\label{OrbitalShiftPPN}
\psi^{PPN}_{prec}=\frac{6\pi G_{N}M}{L}\left(1+\frac{2\gamma-\widetilde{\beta}-1}{3}\right),
\end{equation}
where $\widetilde{\beta}$ and $\gamma$ are Eddington parameters. For the perihelion shift of Mercury, the constraint on PPN parameters provided by Messenger spacecraft \cite{Verma2014} is given by $|2\gamma-\widetilde{\beta}-1|<7.8\times10^{-5}$. Therefore, we obtain
\begin{equation}
\label{MercuryPPNConstr}
\frac{12G_{N}^{2}M^{2}}{c^{4}L_{*}^{2}}=|2\ensuremath{\gamma}-\ensuremath{\widetilde{\beta}}-1|<7.8\times10^{-5},
\end{equation}
which approximately gives
\begin{equation}
\label{MPBound}
5.8\times10^{5}m\lesssim L_{*}.
\end{equation}

On the other hand, the S2 star orbiting around Sagittarius A* gives a laboratory to test general relativity in the strong gravitational field. In our case, it can be provide a much larger lower bound for $L_{*}$. Recently, the GRAVITY Collaboration \cite{Abuter2020} measured the precession of S2's orbit $(2+2\gamma-\widetilde{\beta})/3=1.10\pm0.19$ which gives $1.29$ and $0.91$. Since $\alpha=-1$ for minimum value $0.91$, we only consider maximum value $1.29$. So, we get
\begin{equation}
\label{S2Bound}
\frac{4G^{2}_{N}M^{2}}{c^{4}L_{*}^{2}}<0.29.
\end{equation}
Taking $M=4.25\times10^{6}M_{\odot}, $ the lower bound on $L_{*}$ is given by
\begin{equation}
\label{S2Bound2}
4\times10^{10}m\lesssim L_{*}.
\end{equation}
 
\section{Discussions and conclusions}
\label{sec4}

\begin{table}[ht]
    \centering
\caption{Lower bounds of new fundamental length scale $L_{*}$}
\label{table1}
\begin{tabular}{cc}
\hline 
Test & $L_{*}$\tabularnewline
\hline 
\hline 
Light deflection \cite{Lu2019}& $\ensuremath{9.1\times10^{5}}$m\tabularnewline
Strong lensing (Sgr A{*}) \cite{Lu2019}& $\ensuremath{2\times10^{10}}$m\tabularnewline
Strong lensing (M87)\cite{Lu2019} & $\ensuremath{3\times10^{13}}$m\tabularnewline
Gravitational redshift & $\ensuremath{9\times10^{-2}}$m\tabularnewline
Geodetic precession & $\ensuremath{2\times10^{-1}}$m\tabularnewline
Shapiro time delay & $\ensuremath{9\times10^{5}}$m\tabularnewline
Perihelion shift of Mercury's orbit & $\ensuremath{5.8\times10^{5}}$m\tabularnewline
Precession of S2's orbit & $\ensuremath{4\times10^{10}}$m\tabularnewline
\hline 
\end{tabular}
    \label{tab:my_label}
\end{table}
EUP takes into account position uncertainty correction to standard uncertainty principle, and makes quantum effects available at the large distance scale. In this paper, we investigated the observational constraints for the EUP metric. We studied gravitational redshift, geodetic precession, Shapiro time delay, perihelion shift of Mercury and orbit precession of S2 star. Using the results of Solar system and S2 star orbiting around Sgr A*, we obtained the lower bounds of new fundamental length scale $L_{*}$. In table \ref{table1}, we summarized the lower bounds of $L_{*}$ from various observations. 

As can be seen in table \ref{table1}, the bounds from Earth based experiments such as gravitational redshift and geodetic precession are the smallest bounds, $10^{-2}-10^{-1}$m. Solar scale observations give much bigger bounds, $10^{5}-10^{6}$m. Beyond the Solar system, the bound $10^{10}$m from the precession of S2 star's orbit is the biggest bound in this work. 
Comparing our bounds with ref. \cite{Lu2019}, the lower bound $10^{13}$m from strong gravitational lensing is the biggest bound for the supermassive black hole in M87.

Before finishing the paper, we give some comments on the nature of $L_{*}$. One may ask whether $L_{*}$ is universal just like its counterpart Planck length $L_{Pl}$ or depends on a particular gravitational system. Although  $L_{*}$ does not have a well defined value, one may expect that $L_{*}$ has one value. In order to affect the physics of supermassive black holes, the value of $L_{*}$ must be sufficiently large in this case ($L_{*}\sim10^{10}m$ or beyond). In the second case, one may consider $L_{*}$ depending on the the mass of a particular gravitational system. In this case, $L_{*}$ varies between $10^{-2}-10^{13} m$ according to this work and ref.\cite{Lu2019}. However, the second case may not be favourable, because it is well-known that the Solar system tests are not sensitive tools to set precise bounds on the large scale structures \cite{Kagramanova2006}.

The observational constraints for EUP may open a new window to understand the quantum features at large distance scale. Since new fundamental length scale $L_{*}$ may play a key role in the properties of supermassive black hole, more research is needed in the future.

\section*{Acknowledgments}
	The authors thank the anonymous referee for his/her helpful and constructive comments. Özgür Ökcü thanks Christian Corda for reading the manuscript and the fruitful discussion. This work was supported by Istanbul University Post-Doctoral Research Project: MAB-2021-38032.
		
\section*{Data availability}
		No new data were created or analysed in this study.

\end{document}